# Correction to Euler's equations and elimination of the closure problem in turbulence.

Michail Zak
Senior Research Scientist (Emeritus)
Jet Propulsion Laboratory California Institute of Technology
Pasadena, CA 91109

**Abstract.**

It has been demonstrated that the Euler equations of inviscid fluid are incomplete: according to the principle of release of constraints, absence of shear stresses must be compensated by additional degrees of freedom, and that leads to a Reynolds-type multivalued velocity field. However, unlike the Reynolds equations, the enlarged Euler's (EE) model provides additional equations for fluctuations, and that eliminates the closure problem. Therefore the EE equations are applicable to fully developed turbulent motions where the physical viscosity is vanishingly small compare to the turbulent viscosity, as well as to superfluids and atomized fluids. Analysis of coupled mean/fluctuation EE equations shows that fluctuations stabilize the whole system generating elastic shear waves and increasing speed of sound. Those turbulent motions that originated from instability of underlying laminar motions can be described by the modified Euler's equation with the closure provided by the stabilization principle: driven by instability of laminar motion, fluctuations grow until the new state attains a neutral stability in the enlarged (multivalued) class of functions, and these fluctuations can be taken as boundary conditions for the EE model. The approach is illustrated by an example.

## Introduction

In Newtonian mechanics, the models of continua are derived from the Newton's laws. However in addition to that, several *mathematical restrictions*, (such as space-time differentiability, the Lipchitz conditions etc.) which are not required by the Newton's laws, are exploited. Some of these restrictions have a good physical justification: loss of differentiability could lead to unbounded stresses, and that is not acceptable in physical models. Indeed in classical continua, and in particular, in the model of a fluid, the stress tensor depends upon the tensor of the velocity of deformation as following

$$\sigma_{ik} = -p\delta_{ik} + \eta(\frac{\partial v_i}{\partial x_k} + \frac{\partial v_k}{\partial x_i}) \tag{1}$$

where $\sigma_{ik}$ stress, $p$ pressure, $v$ velocity, $x$ space coordinate, and $\eta$ viscosity. Therefore

$$\sigma \to \infty \ if \qquad \partial v / \partial x \to \infty \tag{2}$$

However the model of inviscid fluid is an exception: there by definition

$$\eta \equiv 0 \tag{3}$$

and, as follows from Eq. (1), non-differential velocity field does not effect stresses. It means that requirement of differentiability of the velocity field in the model of inviscid fluid is introduced only for mathematical convenience, and it can be removed. There are many alternatives to differential velocity field, and we will choose here the simplest one. A hint for our choice follows from the properties of the velocity field in turbulent motions:

$$v(x_1) \neq v(x_2) \quad if \quad x_1 \rightarrow x_2 \tag{4}$$

i.e. very close points can have very different velocities. We will move this property to the extreme assuming that

$$v(x_1) \neq v(x_2) \quad if \quad x_1 = x_2 \tag{5}$$

Such an idealization means that two different particles with different velocities can appear at the same point of space without causing any physical inconsistency, and that is possible only due to the condition (3). Actually we arrive at two superimposed continua, and each of them can be described by the corresponding Euler equations

$$\frac{\partial v_1}{\partial t} + v_1 \nabla v_1 = \frac{1}{2\rho} \nabla p + F \tag{6}$$

$$\frac{\partial v_2}{\partial t} + v_2 \nabla v_2 = \frac{1}{2\rho} \nabla p + F \tag{7}$$

$$\nabla \bullet (v_1 + v_2) = 0 \tag{8}$$

where $F$ is external force per unit mass.

It should be emphasized that the pressure $p$, the density $\rho$ as well as the divergence of velocity must remain single-valued. Indeed as follows from Eq. (1), only the deviatoric components of the tensor of velocity deformation are "released" from the requirement of differentiability, while $p$, $\rho$ and $\nabla \bullet (v)$ do not depend on these components.

For further transformations, the following decomposition of the velocity field will be useful

$$v = \bar{v} \pm \tilde{v}, \qquad \bar{v} = \frac{1}{2}(v_1 + v_2), \qquad \tilde{v} = \pm \frac{1}{2}(v_1 - v_2) \tag{9}$$

Here $\bar{v}$ is the velocity of the "center of inertia" of the two particles superimposed at the same point of space (an analog of the classical velocity), and $\pm \tilde{v}$ are the fluctuation with respect to the "center of inertia" introduced above.

Adding up Eqs.(6) and (7), and subtracting them from one another one obtains respectively

$$\frac{\partial \bar{v}}{\partial t} + \bar{v} \nabla \bar{v} + \tilde{v} \nabla \tilde{v} = \frac{1}{\rho} \nabla p + F \tag{10}$$

$$\frac{\partial \tilde{v}}{\partial t} + \bar{v} \nabla \tilde{v} + \tilde{v} \nabla \bar{v} = 0 \tag{11}$$

Eq. (8) in new variables can be rewritten in the form

$$\nabla \bullet (\bar{v}) = 0 \tag{12}$$

For compressible inviscid fluid, Eq. (12) should be replaced by the following

$$\frac{\partial \rho}{\partial t} + \nabla \bullet (\rho v) = 0, \qquad p = f(\rho) \tag{13}.$$

For an atomized fluid when $p \equiv 0$, following the argumentations discussed above, one should remove the requirement of differentiability of the divergence $\nabla \bullet v$ and arrive at the following EE governing equations

$$\frac{\partial \bar{v}}{\partial t} + \bar{v} \nabla \bar{v} + \tilde{v} \nabla \tilde{v} = F \tag{14}$$

$$\frac{\partial \widetilde{v}}{\partial t} + \overline{v}\nabla\widetilde{v} + \widetilde{v}\nabla\overline{v} = 0 \qquad (15)$$

The system (10) (11) and (13), as well as the system (14) and (15) represent the enlarged Euler's (EE) equations of inviscid fluid. This system slightly resembles the Reynolds equations, [1], but there are several fundamental differences listed below.

1. The EE systems are closed, i.e. the number of unknowns is equal to the number of equations, and that eliminates the closure problem.

2. The EE systems do not include the continuity equation for fluctuations.

These differences follow from the fact that the Reynolds velocity field, strictly speaking, is single-valued since the stress tensor of the Navier-Stokes equations does not have zero components. In other words, the condition (5) for the Reynolds velocity field should be replaced by a weaker condition (4).

  The double-valued EE model introduced above can be generalized to *n*-valued EE model with the same physical justification. In this case an inviscid fluid is a result of superposition of *n* physically identical, but kinematically different continua. In case of incompressible fluid there are *n+1* governing equations with respect to *n+1* independent variables

$$\frac{\partial v_i}{\partial t} + v_i\nabla v_i = \frac{1}{n\rho}(\nabla p + F), i = 1, 2, \dots n.$$ (16)

coupled via the mass conservation equation

$$\nabla \bullet \sum_{i=1}^{n} v_i = 0 \qquad (17)$$

However for the sake of clarity, in this paper we will concentrate only on the double-valued model.

## 2. General characteristics of EE equations.

*a. Bernoulli integrals*. In this section we will investigate properties of the EE system that can be formulated without invoking specific initial and boundary conditions such as first integrals, conservations laws, and characteristic speed of wave's propagation. For that purpose let us present Eqs. (10), (11) and (13) in a more convenient form

$$\frac{\partial \overline{v}}{\partial t} + \nabla \bullet (\frac{1}{2}\overline{v}^2 + \frac{1}{2}\widetilde{v}^2) - \overline{v}\times(\nabla\times\overline{v}) - \widetilde{v}\times(\nabla\times\widetilde{v}) = F - \frac{1}{\rho}\nabla p \qquad (18)$$

$$\frac{\partial \widetilde{v}}{\partial t} + \nabla \bullet (\frac{1}{2}\overline{v}^2 + \widetilde{v}\bullet\overline{v}) - \overline{v}\times(\nabla\times\widetilde{v}) - \widetilde{v}\times(\nabla\times\overline{v}) = 0 \qquad (19)$$

For a stationary potential motion we have an EE analog of the Bernoulli integral

$$\frac{\overline{v}^2}{2} + \frac{\widetilde{v}^2}{2} + \int_{p_0}^{p}\frac{dp}{\rho(p)} + \Pi = const., \qquad -\nabla\bullet\Pi = F \qquad (20)$$

This integral reduces to classical one when $\widetilde{v} = 0$.

But there is another integral that does not have a classical analog

$$\frac{\widetilde{v}^2}{2} + \widetilde{v}\bullet\overline{v} = const. \qquad (21)$$

and that vanishes when $\widetilde{v} = 0$.

*b. Vortex lines conservation*. Applying the operator $(\nabla\times)$ to Eqs. (18) and (19), one finds

$$helm(\nabla \times \overline{v}) = \{\nabla \times f + \frac{1}{\rho^2}(\nabla \bullet \rho) \times (\nabla \bullet p) + \nabla \times [\widetilde{v} \times (\nabla \times \widetilde{v})] \tag{22}$$

$$helm(\nabla \times \widetilde{v}) = \nabla \times [\overline{v} \times (\nabla \times \widetilde{v})] + \nabla \times [(\widetilde{v} \times (\nabla \times \overline{v})] \tag{23}$$

As follows from Eq. (22), for conservation of vortex lines of the mean field, the conditions formulated by Helmholtz should be complemented by an additional constraint imposed upon the fluctuations

$$\nabla \times [\widetilde{v} \times (\nabla \times \widetilde{v})] \equiv 0 \tag{24}$$

The condition for conservation of the vortex lines of the fluctuation field follows from Eq. (23)

$$\overline{v} \equiv 0 \tag{25}$$

***c. Speed of sound***. Let us find a speed of sound $a$ in EE model of a compressible fluid. As follows from Eqs. (10), (11) and (13), for jumps of the derivatives $[\partial \overline{v}/\partial n]$ and $[\partial \widetilde{v}/\partial n]$ projected onto the normal $n$ to the surface of discontinuity

$$(\overline{v}_n - \lambda - \frac{\partial p/d\rho}{\overline{v}_n - \lambda})[\partial \overline{v}/\partial n] + \widetilde{v}_n[\partial \widetilde{v}/\partial n] = 0 \tag{26}$$

$$(\overline{v}_n - \lambda)[\partial \widetilde{v}/\partial n] + \widetilde{v}_n[\partial \overline{v}/\partial n] = 0 \tag{27}$$

whence

$$a_{EE} = \lambda - \overline{v}_n = \pm\sqrt{\frac{dp}{d\rho} + \widetilde{v}_n{}^2} \qquad if \qquad [\partial \overline{v}/\partial n] \neq 0 \tag{28}$$

while $\lambda$ is the characteristic speed of the wave propagation that is supposed to satisfy the kinematical compatibility condition at the front of the discontinuity

$$[\frac{\partial}{\partial t}] = -\lambda[\frac{\partial}{\partial n}] \tag{29}$$

Thus the speed of sound in an EE flow is larger than that in a laminar flow, and the difference increases with increase of the kinetic energy of fluctuations. There is another interesting observation that follows from Eq. (28): since fluctuation projections on different normal directions could be different, the speed of sound in an EE medium depends upon the direction of sound.

In an atomized fluid, the sound propagates with the speed

$$a_{EE} = \lambda - \overline{v}_n = \pm\widetilde{v}_n \tag{30}$$

i.e. by convection of fluctuations.

***d. Mach's angle***. Let us investigate stationary discontinuities in a potential EE motion and find the angle $\alpha$ between the mean velocity $\overline{v}$ and the tangent $\tau$ to the line of discontinuity, (the Mach's angle). Introducing the direction $n$ orthogonal to the line of discontinuity and turning to Eqs. (10), (11) and (13), one obtains

$$\overline{v}_n[\partial \overline{v}/\partial n] + \widetilde{v}_n[\partial \widetilde{v}/\partial n] + \frac{1}{\rho}\frac{dp}{d\rho}[\rho] = 0 \tag{31}$$

$$\overline{v}_n[\partial \widetilde{v}/\partial n] + \widetilde{v}_n[\partial \overline{v}/\partial n] = 0 \tag{32}$$

$$\rho[\partial \overline{v}/\partial n] + \overline{v}_n[\partial \rho/\partial n] = 0 \tag{33}$$

whence

$$\overline{v}_n = \overline{v}\sin\alpha = \pm\sqrt{dp/d\rho + \widetilde{v}_n{}^2} \quad and \quad \sin\alpha = \pm a_{EE}/\overline{v} \tag{34}$$

Therefore the Mach's angle in EE model is larger than the classical one.

***e. Elastic shear waves.*** Let us demonstrate that in EE model there exists a new type of elastic shear waves similar to those in elastic bodies. For that purpose, we project Eqs. (10) and (11) onto the normal $n$ to the surface of tangential discontinuity of the velocities (with the tangent $\tau$). Taking into account that, as follows from Eq. (13)

$$[\partial \tilde{v}_\tau / \partial \tau] = 0, \tag{35}$$

and as follows from the conditions of kinematical compatibility at the surface of discontinuity,

$$[\partial / \partial n] = 0 \qquad if \qquad [\partial / \partial \tau] \neq 0 \tag{36}$$

one obtains

$$(\bar{v}_\tau - \lambda)[\partial \bar{v}_n / \partial \tau] + \bar{v}_\tau [\partial \tilde{v}_n / \partial \tau] = 0 \tag{37}$$

$$(\bar{v}_\tau - \lambda)[\partial \tilde{v}_n / \partial \tau] + \tilde{v}_\tau [\partial \tilde{v} / \partial \tau] = 0 \tag{38}$$

whence

$$\lambda = \bar{v}_\tau \pm \tilde{v}_\tau \tag{39}$$

By analogy to shear waves in elastic bodies, one can introduce an equivalent of the shear modulus

$$G = \rho \tilde{v}_\tau^2 \tag{40}$$

It should be emphasized that the result (40) is valid for both compressible and incompressible EE continua, but the necessary condition for existence of elastic shear waves is a rotational motion. Otherwise the equalities

$$[\partial \bar{v}_n / \partial \tau] = 0, [\partial \tilde{v}_n / \partial \tau] = 0 \tag{41}$$

would follow from the conditions

$$[\partial \bar{v}_\tau / \partial n] = 0, [\partial \tilde{v}_\tau / \partial n] = 0 \tag{42}$$

that hold in potential motions.

It means that actually the shear waves transport vortices $\nabla \times v$.

***f. EE model of turbulence***. Let us compare motions described by the EE equations to turbulence, as we know it. A developed turbulence is characterized by mean and fluctuation velocities, while fluctuations can be divided in two classes: small and large scale fluctuations. The small-scale fluctuations ($\mathrm{Re} \approx 1$) are responsible for dissipation of mechanical energy, and practically they do not affect the general picture of motion since their amplitudes are small compare to mean velocities. The large-scale fluctuations ($\mathrm{Re} \to \infty$) are sizable with mean velocities, and they significantly contribute to the motions. These properties suggest that the general picture of turbulence is better captured by the Euler rather than Navier-Stokes equation, and in particular, by EE equations, regardless of whether the underlying pre-instability laminar flow is viscous or non-viscous. Now the problem is to describe the transition to EE from unstable laminar flows. A hint for that follows from the stabilizing properties of fluctuations making EE model more resistible to shear deformations, and therefore, more stable than the Euler or the Navier-Stokes equations. For the proof of concept we will turn to an analytical example.

## 3. Transition from laminar flows to EE turbulence. Example.

Let us consider a surface of a tangential jump of velocity $\bar{V}_2 - \bar{V}_1$ in a horizontal unidirectional flow of an inviscid incompressible fluid assuming that this surface is not penetrated by the mean velocity $\bar{V}$ of the double-valued velocity field. However we will allow this surface to be penetrated by fluctuations of this field. Actually we introduced the following constraint

$$\overline{V}^n = 0 \tag{43}$$

where $n$ is the normal to the surface of discontinuity, Fig.1.

In order to investigate the stability of the flow we will analyze the speed of propagation of high frequencies of the surface shape oscillations that are equal to characteristic speeds of propagation of discontinuities of the corresponding derivatives.

Applying the principle of virtual work to a small volume $V$ containing both flows of the fluid as well as the surface of the tangential jump of velocities separating the flows one obtains

$$\int_V (\tilde{\rho} a_1 \bullet \delta U_1 + \tilde{\rho} a_2 \bullet \delta U_2) dV = 0 \tag{44}$$

where $\tilde{\rho}$, $U_1$, $U_2$, $a_1, a_2$ are density, displacements, and accelerations of the fluid. The displacements $U_1$ and $U_2$ are mutually independent in the region that does not contain a separating surface, but they are dependent at the surface due to its mean field velocity impenetrability

$$U_1 \bullet n = U_2 \bullet n = U, \tag{45}$$

Hence, as follows from Eq. (44), at the surface the following equality holds:

$$(a_1 + a_2) \bullet n = 0, i.e. \qquad \{[\frac{\partial}{\partial t} + (\overline{V}_1 + \tilde{V})\frac{\partial}{\partial S}]^2 + [\frac{\partial}{\partial t} + (\overline{V}_2 - \tilde{V})\frac{\partial}{\partial S}]^2\}U + \alpha = 0 \tag{46}$$

where S is a coordinate along the line of discontinuity, and $\alpha$ is the term that does not contain second order derivatives of $U$, and therefore, it does not effect the characteristic equation

$$(\lambda - \overline{V}_1 - \tilde{V})^2 + (\lambda - \overline{V}_2 + \tilde{V})^2 = 0 \tag{47}$$

and its characteristic roots

$$\lambda = \frac{1}{2}[(\overline{V}_2 + \overline{V}_1) \pm i(\overline{V}_2 - \overline{V}_1 - 2\tilde{V})] \tag{48}$$

In the classical case when

$$\tilde{V} = 0, \text{ and } \overline{V} = V \tag{49}$$

we arrive at the well known result

$$\lambda' = \frac{1}{2}[(V_2 + V_1) \pm i(\overline{V}_2 - \overline{V}_1)]. \tag{50}$$

Let us start with studying propagation of high frequency oscillations of the transverse displacements $U$. Recall that Eq. (46) being linear with respect to the second order time-space derivatives, strictly speaking, is nonlinear with respect to $U$ and its first time-space derivatives that are contained in the term $\alpha$. For small amplitudes and their first derivatives, this term can be linearized:

$$\alpha = \alpha_1 \frac{\partial U}{\partial t} + \alpha_2 \frac{\partial U}{\partial S} + \alpha_3 U + \alpha_4 \qquad (51)$$

For further simplifications, all the coefficients in Eq.(46) can be linearized with respect to an arbitrarily selected point $S_0$ and instant of time $t_0 = 0$. Then Eq. (46) takes form of a linear elliptic PDE with constant coefficients

$$[(\frac{\partial}{\partial t} + V_1^{\ 0} \frac{\partial}{\partial S})^2 + (\frac{\partial}{\partial t} + V_2^{\ 0} \frac{\partial}{\partial S})^2 + \alpha_1^0 \frac{\partial}{\partial t} + \alpha_2^0 \frac{\partial}{\partial S} + \alpha_3^0]U + \alpha_4^0 = 0 \qquad (52)$$

Obviously this equation is valid only for small amplitudes with small first derivatives, the small area around the above selected point $S_0$, and within a small period of time $\Delta t$.

Let us derive the solution to Eq. (52) subject to the following initial conditions

$$U*_0 = \frac{1}{\lambda_0} e^{-\lambda_0 Si}, \ at \qquad t = 0 \qquad (53)$$

assuming that $\lambda_0$ can be made as large as desired, i.e. $\lambda_0 > N \to \infty$. Consequently, the initial disturbances can be made as small as desired, i.e. $U*_0 < N^{-1} \to 0$. The corresponding solution can be written in the form

$$U* = C_1 e^{-\lambda_0 (\lambda_1' t - S)i} + C_2 e^{-\lambda_0 (\lambda_2' t - S)i} \qquad (54)$$

where $\lambda_1', \lambda_2'$ are the roots of the characteristic equation (50) . Since the characteristic roots are complex, the solution (54) will contain the term

$$\frac{1}{\lambda_0} e^{|\mathrm{Im}\,\lambda_{1,2}'|\Delta t} \sin \lambda_0 S, \qquad \lambda_0 \to \infty \qquad (55)$$

that leads to infinity within an arbitrarily short period of time $\Delta t$ and within an infinitesimal area around the point $S_0$ .

Hence one arrived at the following situation: $|U*| \to \infty$ in spite the fact that $|U_0| \to 0$. In order to obtain a geometrical interpretation of the above described instability, let's note that if the second derivatives in Eq. (52) are of order $\lambda_0$, then the first derivatives are of order of $I$, and $U$ is of order of $I/\lambda_0$. Hence, the period of time $\Delta t$ can be selected in such a way that the second derivatives will be as large as desired, but $U$ and its first derivatives are still sufficiently small. Taking into account that the original governing equation (46) is quasi-linear with respect to the second derivatives, and therefore, the linearization does not impose any restrictions on their values, one concludes that the linearized equation (52) is valid for the solution during the above mentioned period of time $\Delta t$. Turning to the term (55) of the solution (54), one can now interpret it as being represented by a function having an infinitesimal amplitude and changing its sign with an infinite frequency ($\lambda_0 \to \infty$). The first derivatives of this function can be small and change their signs by finite jumps (with the same infinite frequency), so that the second derivatives at the points of such jumps are infinite. From the mathematical viewpoint, this kind of function is considered as continuous, but *non-differentiable*.

The result formulated above was obtained under specially selected initial conditions (53), but it can be generalized to include any initial conditions. Indeed, let the initial conditions be defined as

$$U\mid_{t=0} = U**_0(X) \tag{56}$$

and the corresponding solution of Eq. (46) is

$$U = U**(X, t) \tag{57}$$

Then, by altering the initial conditions to

$$U\mid_{t=0} = U_0*(X) + U_0**(X) \tag{58}$$

where $U*_0$ is defined by Eq. (53), one observes the preceding argument by superposition that vanishingly small change in the initial condition (57) leads to unboundedly large change in the solution (56) that occurs during an infinitesimal period of time. That makes the solution (56) *non-differentiable*, but still continuous. However, the Euler's equation that governs the flow discussed above was derived under condition *of space-time differentiability* of the velocity field. This discrepancy manifests inapplicability of the Euler's equation for description of *postinstability* motion of an inviscid fluid, and therefore, for a developed turbulence. But this does not imply the incompleteness of Newtonian mechanics: it only means that the mathematical formalism that expresses the Newton's laws should include non-differentiable components of the velocity field

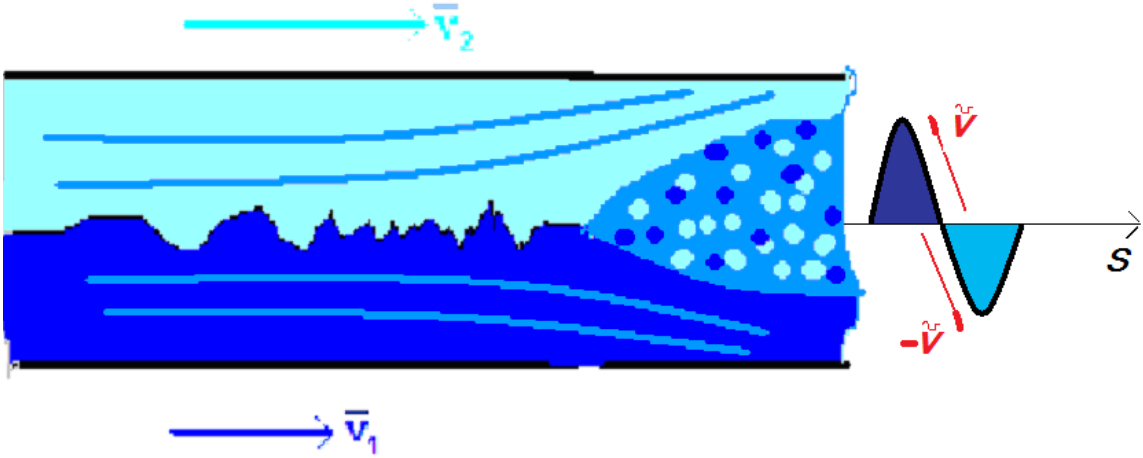

**Figure 1. Loss of differentiability in fluid mechanics.**

Let us try to solve the same problem by introducing the EE model and turn to the general form of the characteristic equation (47) where the velocity fluctuations are present.

The solution of the EE version of the same equation has the form (54) in which the characteristic roots $\lambda_1, \lambda_2$ are defined by Eq. (48) instead of Eq. (50)

$$U* = C_1 e^{-\lambda_0(\lambda_1 t - S)i} + C_2 e^{-\lambda_0(\lambda_2 t - S)i} \tag{59}$$

As long as the characteristic roots (50) have non-zero imaginary parts, the solution (59) has the property similar to those of the solution (54) discussed above.

Now we can describe the scenario of transition from unstable laminar flow to EE model of turbulence. We will start with the assumption that even in a laminar flow there always exist infinitesimal fluctuations of the fluid velocities. Then, driven by the mechanism of instability, these fluctuations grow until they stabilize the flow, and that represents the stabilization principle formulated, discussed, and illustrated in [2-6]. Turning to the characteristic roots (48) one can see that the fluctuations that eliminate the imaginary part of these roots must be

$$\tilde{v} = \pm \frac{1}{2}(\bar{V}_1 - \bar{V}_2) \qquad (60)$$

Then the characteristic equation (48) is modified to

$$(\lambda - \bar{V}_1 + \tilde{v})^2 + (\lambda - \bar{V}_2 - \tilde{v})^2 = 0 \qquad (61)$$

and the double characteristic root is real

$$\lambda_1 = \lambda_2 = \frac{1}{2}(\bar{V}_1 + \bar{V}_2) \qquad (62)$$

Thus the shape of the line of discontinuity $S$ becomes neutrally stable due to fluctuations that being driven by the mechanism of instability grew up to the magnitude (60).

We will make the following comment to this example.

Although we have solved a linearized version of the problem formulated above, nevertheless this solution is sufficient for formulating the conditions of instability and describing the transition from unstable laminar flow to EE turbulence. And the explanation of that comes from the fact that we are dealing hear with the Hadamard (or blow-up) instability rather than the Lyapunov instability; the Hadamard instability is caused by transition from a hyperbolic to an elliptic type of the corresponding PDE, and that transition involves only higher order derivatives. But since the PDE (46) is *linear with respect to the higher order derivatives,* the linearization leads to exact results. However, in order to solve the problem in full, one has to invoke Eqs. (10), (11) and (12) that describe the velocity field around the surface on discontinuity while the found velocities at this surface have to play the role of boundary condition to these equations.

Let us now *formulate* the whole problem for the example considered above. We have to start with the equations (10), (11) and (12)

$$\frac{\partial \bar{v}}{\partial t} + \bar{v} \nabla \bar{v} + \tilde{v} \nabla \tilde{v} = \frac{1}{\rho} \nabla p + F$$

$$\frac{\partial \tilde{v}}{\partial t} + \bar{v} \nabla \tilde{v} + \tilde{v} \nabla \bar{v} = 0$$

$$\nabla \bullet (\bar{v}) = 0$$

These two vectors and one scalar equations contain two unknown vectors $\bar{v}$ and $\tilde{v}$, and one unknown scalar $p$, $\rho$ and therefore, the system is closed. The system should be solved subject to the following initial and boundary conditions

*The initial conditions at t=0*

$\overline{v} = \overline{V}_1$ -   under the surface of discontinuity           (63)

$\overline{v} = \overline{V}_2$ -   above the surface of discontinuity           (64)

*The boundary conditions at the lower part of the surface of discontinuity*

$$\overline{v}^{\tau} = \overline{V}_1, \overline{v}^n = 0$$

$$\tilde{v}^{\tau} = \frac{1}{2}(\overline{V}_2 - \overline{V}_1), \tilde{v}^n = 0 \qquad (65)$$

*The boundary conditions at the upper part of the surface of discontinuity*

$$\overline{v}^{\tau} = \overline{V}_2, \overline{v}^n = 0$$

$$\tilde{v}^{\tau} = \frac{1}{2}(\overline{V}_1 - \overline{V}_2), \tilde{v}^n = 0 \qquad (66)$$

In addition to that, *at the surface of discontinuity*

$$p^- = p^+ \quad p^- = p^+ \qquad (67)$$

i.e. the pressure remains continuous at the surface of *velocity* discontinuity.

Since the boundary conditions (65) and (66) include non-zero fluctuations (found from the stability analysis of the shape of the surface of discontinuity), the fluctuations are generated at all the space under consideration, and that represents the EE model of turbulence.

Although this result has a methodological value illustrating the stabilization principle as a bridge between pre-instability laminar motion and post-instability turbulent motion represented by the EE model, its practical significance is limited: strictly speaking, it can be applied only to superfluids (liquid helium, and some of Bose-Einstein condensates), i.e. fluids with no viscosity at all. Indeed in fluids with non-zero viscosity, no matter how small it is, a finite tangential jump of velocity would cause an unbounded shear stress. That is why for classical fluids no-slip condition at surfaces of tangential jumps of velocities (including rigid boundaries) must be enforced, and the EE model should be applied only beyond the corresponding boundary layer, while a connection between the laminar motion within this layer and the turbulent motion beyond it is to be implemented by the stabilization principle as illustrated above. In addition to that, the EE model can be applied to atomized fluid (see Eqs.(14) and (15)) with application to the cavitation processes. However, a major limitation of this approach is a necessity to find the rate of instability of the original laminar flow prior to application of the stabilization principle, and this pre-condition is very complex and laborious since the criteria of instability of the Navier-Stokes equations can be analytically found in a very limited number of cases. Therefore at this stage only numerical approach to implementation of the EE model seems visible. A possible computational strategy that allows one to compute mean velocity as well as higher moments based upon the stabilization principle is proposed in [4-6].

However if we are not interested in an onset of turbulence and are dealing with a stationary turbulence, the problem becomes simpler: the stationary version Eqs.(10-12)

$$\overline{v}\nabla\overline{v} + \tilde{v}\nabla\tilde{v} = \frac{1}{\rho}\nabla p + F \qquad (68)$$

$$\overline{v}\nabla\tilde{v} + \tilde{v}\nabla\overline{v} = 0 \tag{69}$$

$$\nabla \bullet (\overline{v}) = 0 \tag{70}$$

must be solved subject to known boundary conditions. Obviously for that solution *no specific* computational strategy is needed.

## 4. Discussion and conclusion.

The main objective of this paper is to modify the Euler's equations and to **formulate** the governing equations for turbulent motion. Obviously that must be done **prior** to developing any computational strategy. That is why for a proof of concept, our attention here was concentrated on demonstration and discussion of an **analytical** example rather than on a general computational strategy. At the same time, a computational strategy that may be applicable for the EE formulation of the problem of turbulence has been proposed in our previous publications, [4-6], and it can complement the results of this paper.

The main "shift of paradigm "in this paper is a departure from the requirement of differentiability of the velocity field in the model of an inviscid fluid. The rationale for this departure is that in case of inviscid fluid, zero shear stress must be compensated by additional degree of freedom in accordance to the principle of release of constraint, [7]. This principle associates any reaction to a constraint with a "lost" degree of freedom, and thereby it allows applying the second Newton's law to a constrained point as to a free point with a reaction of the constraint as an additional force. Then the inverse is true: a removal of the constraint releases the additional degree of freedom associated with this constraint, and that degree of freedom, in case of inviscid fluid, can be implemented via double-valued velocity field. Hence it is not a coincidence that only in case of zero shear stress the multivaluedness of deviatoric parts of the velocity deformation tensor does not lead to unbounded components of stress tensor, and this allows one to introduce the EE model. One of the basic results of the EE characteristics analysis is that EE model is more stable than the underlying Euler or Navier-Stokes equations describing laminar flows, while the degree of the stability is proportional to the kinetic energy of fluctuations. Actually this supports the stabilization principle i.e. transition from unstable laminar motion to stable EE motion.

In this context, it is instructive to take another look on the concept of instability in Newtonian dynamics. Usually inability of the Newton's laws to discriminate between stable and unstable motions is considered as a fundamental limitation of classical mechanics. However as follows from the stabilization principle, *stability is not a physical invariant*. That is why it requires an additional *mathematical* analysis.

Thus it has been demonstrated that the Euler equations of inviscid fluid are incomplete: according to the principle of release of constraints, absence of shear stresses must be compensated by additional degrees of freedom, and that leads to a Reynolds-type multivalued velocity field. However, unlike the Reynolds equations, the enlarged Euler's (EE) model provides additional equations for fluctuations, and that eliminates the closure problem. Therefore the EE equations are applicable to fully developed turbulent motions where the physical viscosity is vanishingly small compare to the turbulent viscosity, as well as to superfluids and atomized fluids. Analysis of coupled mean/fluctuation EE equations shows that fluctuations stabilize the whole system generating elastic shear waves and increasing speed of sound. Those turbulent motions that originated from instability of underlying laminar motions can be described by the modified Euler's equation with the closure provided by the stabilization principle: driven by instability of laminar motion, fluctuations grow until the new state attains a neutral stability in the enlarged (multivalued) class of functions, and these fluctuations can be taken as boundary conditions for the EE model. The approach is illustrated by an example.

**References.**